\documentclass{article}

\usepackage{graphicx} % Required for inserting images
\usepackage{geometry}
\usepackage[colorlinks,bookmarksopen,bookmarksnumbered,citecolor=red,urlcolor=red]{hyperref}

\usepackage{amsmath}
\usepackage{amssymb}
\usepackage{amsfonts}

\title{\textbf{A Novel Convolutional-Free Method for 3D Medical Imaging Segmentation}}
\author{\textbf{Canxuan Gang} \\AI Geeks\\\url{https://aigeeksgroup.github.io}}
\date{}

\begin{document}
\setlength{\parindent}{0pt}
\setlength{\parskip}{10pt}

\maketitle

% Your content here

\begin{abstract}
    Segmentation of 3D medical images is a critical task for accurate diagnosis and treatment planning. Convolutional neural networks (CNNs) have dominated the field, achieving significant success in 3D medical image segmentation. However, CNNs struggle with capturing long-range dependencies and global context, limiting their performance, particularly for fine and complex structures. Recent transformer-based models, such as TransUNet and nnFormer, have demonstrated promise in addressing these limitations, though they still rely on hybrid CNN-transformer architectures. This paper introduces a novel, fully convolutional-free model based on transformer architecture and self-attention mechanisms for 3D medical image segmentation. Our approach focuses on improving multi-semantic segmentation accuracy and addressing domain adaptation challenges between thick and thin slice CT images. We propose a joint loss function that facilitates effective segmentation of thin slices based on thick slice annotations, overcoming limitations in dataset availability. Furthermore, we present a benchmark dataset for multi-semantic segmentation on thin slices, addressing a gap in current medical imaging research. Our experiments demonstrate the superiority of the proposed model over traditional and hybrid architectures, offering new insights into the future of convolution-free medical image segmentation.
\end{abstract}

\section{Introduction}

Segmentation is a key task in computer vision and has been a primary research focus for many years \cite{1}. One of the most common applications of 3D segmentation in voxel representation is medical imaging, as the images obtained from main medical imaging modalities such as CT, PET, and MRI are typically in 3D volume \cite{2,wu2023bhsd,zhang2024segreg,tan2024segstitch}. Therefore, accurate 3D segmentation is essential for medical image analysis, including diagnosis and treatment planning.

In the field of computer vision, convolutional neural networks (CNNs) have been highly successful in recent decades \cite{3,4,5,6}. Their ability to automatically learn and extract features from image data through hierarchical layers of convolutional and pooling operations has revolutionized many computer vision tasks, including object detection \cite{7,zhang2024meddet,cai2024msdet,cai2024medical}, image classification \cite{3,4,5,6,ji2024sine}, and semantic segmentation \cite{8,ge2024esa,tan2024segkan,zhang2025gamed}. The availability of large datasets, such as ImageNet \cite{9}, and advancements in parallel computing and deep learning frameworks have also contributed to the success of CNNs in computer vision. Thus, CNNs have become the main approach for many computer vision tasks due to their ability to capture complex image features through translation equivariance, local sparse connections, and weight sharing \cite{10}. These properties have also made CNNs highly effective for medical imaging segmentation, where accurate delineation of structures and regions of interest is critical for diagnosis and treatment planning \cite{zhang2024deep,hiwase2025can,zhao2024landmark,qi2025projectedex}. Consequently, the state-of-the-art (SOTA) models for medical imaging segmentation are now largely based on CNN architectures, such as U-Net \cite{11} and U-Net++ \cite{12} in 2D imaging, 3D U-Net \cite{13} and nnUNet \cite{14} in 3D imaging, which have been shown to outperform traditional image processing techniques and other types of neural networks in this domain. 

\begin{figure}
    \centering
    \includegraphics[width=0.6\linewidth]{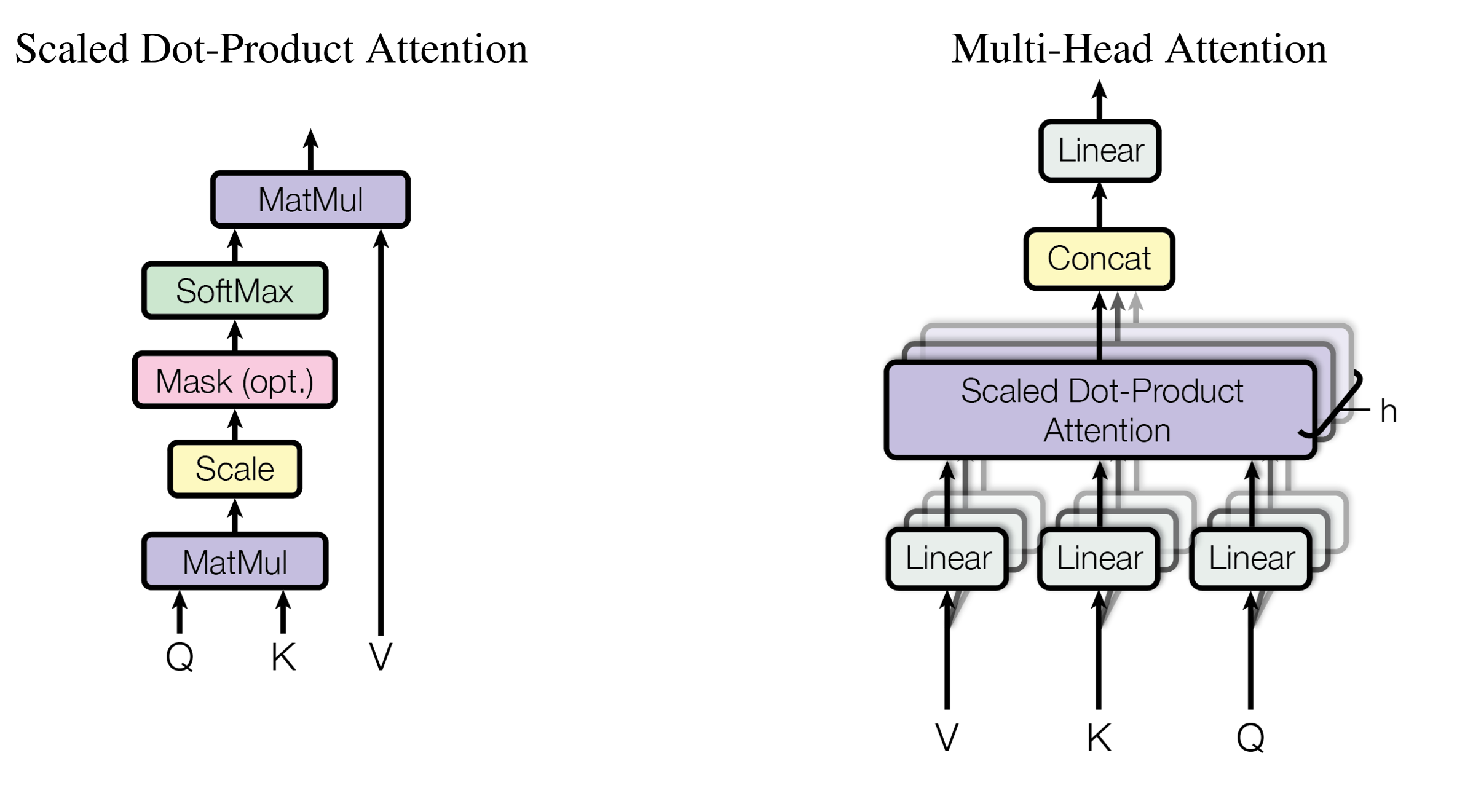} 
    \caption{Self-attention architecture from Transformer \cite{16}}
    \label{fig:1}
\end{figure}

Despite their success in many computer vision tasks, CNNs have been found to have limitations in accurately segmenting fine and complex structures in medical images. Recent research has shown that CNNs can struggle to capture long-range dependencies and global context information \cite{15}, which are critical for accurate segmentation of medical images. In medical imaging segmentation tasks, these limitations are often reflected in low classification accuracy and sub-optimal segmentation quality. To address these limitations, transformer-based architectures with self-attention mechanisms (which shown in figure \ref{fig:1}) have emerged as a promising alternative in general computer vision tasks \cite{17}. For example, the Vision Transformer (ViT) \cite{15} based architecture has demonstrated state-of-the-art performance on various computer vision tasks, including image classification \cite{18}, object detection \cite{19}, and semantic and instance segmentation \cite{20}. With the recent successful implementation of transformer architecture and self-attention mechanism in the general computer vision field, there has been a growing interest in utilizing these techniques for medical imaging segmentation. Several transformer-based models have been proposed, such as TransUNet \cite{21} for 2D segmentation and nnFormer \cite{23} for 3D segmentation. However, while these models have shown promising results, they still rely on combining transformer architecture with convolutional architecture to either extract features or decode masks, making them a compromised architecture rather than entirely "convolutional-free". Thus, the development of a purely convolutional-free architecture based on transformer and self-attention mechanisms is crucial in improving the performance of semantic segmentation and advancing the field of medical imaging segmentation.

Moreover, among various medical imaging and radiological diagnostic modalities, non-contrast computed tomography (NCCT) is one of the most frequently utilized techniques \cite{24}. In the context of medical imaging, there are two types of images that can be obtained from a CT scan: thick slices and thin slices \cite{25}. Thick slices are obtained from thick scans or retrospective reconstruction \cite{26}, whereas thin slices are acquired from thin scans. Unfortunately, publicly available datasets with pixel-wise ground truth are mostly restricted to thick-slice datasets. As a result, existing medical imaging segmentation methods are primarily designed for thick-slice segmentation. However, thick slices are subject to poor resolution in the depth direction, or z-resolution, and are therefore not isomorphic. Conversely, thin slices contain more volumetric information and are isomorphic, meaning they offer superior segmentation results. Additionally, only thin slices can produce a real 3D volume through volumetric rendering (VR) \cite{27}, while interpolation of thick slices leads to the introduction of fake images. This disparity has resulted in a considerable gap in the quality of segmentation outcomes between thin and thick slices. However, hand-annotating a large number of thin slices by a human expert such as a radiologist is not feasible due to the time-consuming nature of the process. Therefore, the development of a novel architecture that incorporates effective strategies for segmenting thin slices based on thick-slice annotations is crucial for the advancement of diagnostic medicine \cite{zhang2023thinthick}.

In general, our research objectives can be summarized as follows: 
\begin{itemize}
    \item To design a novel and efficient model that can significantly enhance the performance of semantic segmentation on 3D medical images, with a focus on improving the classification accuracy of semantic labels and the quality of segmentation masks.
    \item To develop a new architecture that can effectively address the domain adaptation problem between thick and thin slices, allowing for accurate segmentation of thin slices based on thick annotations.
    \item To contribute a new dataset for multi-semantic segmentation on thin slices, addressing the lack of such datasets in the field of medical imaging segmentation.
\end{itemize}
These objectives are critical to advancing the state-of-the-art in medical image analysis and improving diagnostic medicine.

Furthermore, our research approaches can be summarized as follows:

\begin{itemize}
    \item Develop a novel and efficient architecture that is fully convolutional-free and based on transformers to improve the performance of multi-semantic segmentation on 3D medical images. This involves exploring the use of self-attention mechanisms \cite{16} and other techniques to enhance feature extraction and decoding of masks.
    \item Design a joint loss that can supervise the model to segment thin using thick slices. This involves developing a novel loss function that can effectively utilize both thick and thin slices to improve segmentation accuracy.
    \item Publish a benchmark dataset with thin slice multi-semantic segmentation. This involves collecting a large dataset of NCCT images with thin slices and multi-semantic annotations and making it publicly available for researchers to use and evaluate their models.
\end{itemize}

The report will be divided into distinct sections to ensure a coherent flow of information. These sections are as follows:

\begin{itemize}
    \item \textbf{Research Problems}: This section will entail a general description of existing model on 3D medical imaging segmentation. Furthermore, the section will highlight the significance of the proposed method and its potential to address real-world issues. Additionally, it will underscore the benefits of the proposed method when compared to existing models.
    \item \textbf{Methodology}: In this section, the proposed method used to achieve the research objectives will be described. This will involve a detailed explanation of the fully convolutional-free architecture based on transformers, designed to enhance the performance of multi-semantic segmentations. Additionally, the joint loss that supervises the model to segment thin using thick annotations will be presented.
    \item \textbf{Evaluation Criteria}: This section will outline how the model's performance will be evaluated. It will include details on the dataset to be used for evaluation, the design of the experiment for evaluation, and the specific evaluation metrics that will be employed in medical imaging segmentation.
    \item \textbf{Timeline Schedule}: This section will provide a detailed timeline of the research activities and milestones, indicating the expected duration of each activity and the interdependencies between them.
    \item  \textbf{Limitations and Potential Problems}: This section will discuss the potential limitations and challenges of our proposed method, including but not limited to, (1) the potential of utilizing generative models, such as the diffusion model, to address the issue of the long-tail problem by generating synthetic samples of minority classes, (2) the potential capability of large-scale self-supervised pre-training for feature extraction, and (3) potential the ethics and security problem of medicial imaging datasets. We will also explore potential solutions to these problems and discuss future directions for research in this field.
    \item \textbf{Conclusions}: This section will provide a summary of the article, highlighting the key findings and implications for future research. It will also provide recommendations for potential applications of the proposed model in medical imaging segmentation.
\end{itemize}

\section{Research Problems}
\subsection{Related Works}

Prior to introducing our proposed method, we will provide a review of related literature on existing models used in 3D medical imaging segmentation.

The Hough-CNN, proposed by Milletari et al \cite{28} is a typical convolutional neural network designed for 3D medicial imaging segmentations, especially for MRI and Ultrasound. The approach leverages the abstraction capabilities of CNNs and implements Hough voting, which allows for automatic localization and segmentation of anatomies. 

The 3D U-Net, proposed by Çiçek et al \cite{13}, stands out as the first modern model specifically designed for 3D medical imaging segmentations. The architecture of this model is based on the fully-convolutional U-Net proposed by Ronneberger et al. During its inception, the authors acknowledged the unfeasibility and time-consuming nature of labeling a dense ground truth for a 3D volumetric image. As a result, the method employed dense segmentations based on sparse annotations using weighted loss. It should be noted, however, that the lack of publicly available code associated with this work presents a notable limitation. Despite this limitation, the paper's significant influence on subsequent works remains undeniable and has served as a source of inspiration for our proposed method.

The Residual Symmetric 3D U-Net, proposed by Lee et al. \cite{29}, was inspired by the ResNet architecture proposed by Kaiming He et al. \cite{6} in 2016. The key innovation of this approach is the introduction of residual connections between each convolutional layer in the 3D U-Net. This method allows for the successful implementation of residual connections in a fully convolutional architecture, enabling the design of deeper encoder-decoder networks that can extract more abstract features while avoiding issues such as gradient explosion and vanishing. The use of residual connections ensures that the network can learn residual mappings that facilitate the optimization process and improve performance. Overall, the Residual Symmetric 3D U-Net represents an important advancement in the field of 3D medical image segmentation by leveraging the benefits of both residual connections and fully convolutional networks.

The nnU-Net, proposed by Isensee et al \cite{14}. has made significant advancements in the field of 3D medical imaging segmentation, and is widely regarded as the state-of-the-art model in this area. This is due to its innovative pipeline, which builds upon the well-established 3D U-Net architecture. While the backbone of the model remains the same, the authors have introduced novel pre-processing and post-processing techniques, as well as incorporating additional metadata in the widely-used nifti format for CT images. This incorporation of metadata is a significant contribution to the field, as it allows for a more comprehensive understanding of the image and its features. This, in turn, leads to more accurate and reliable segmentation results. The success of nnU-Net has inspired subsequent works, such as CoTr and nnFormer, both of which have also built upon the 3D U-Net architecture. However, nnU-Net remains the most performant model for medical imaging segmentation, due to its ability to leverage metadata and its incorporation of state-of-the-art techniques.

The TransUNet, proposed by Chen et al. \cite{21}, is a pioneering approach to medical imaging segmentation using transformer architecture. The model leverages the transformer layer at the end of the CNN encoder, creating a CNN-Transformer hybrid encoder. This is combined with a cascaded upsampler as a decoder to generate masks. The transformer layer enhances the model's capability to capture global context information, while the CNN encoder still plays a crucial role in feature extraction. However, it is important to note that TransUNet is essentially a 2D segmentation method that concatenates 2D images into 3D, which can result in inconsistencies in masks between scans. This can be observed from the sagittal and coronal views. Despite this limitation, TransUNet has laid the foundation for future research for transformers in medical imaging segmentation.

The CoTr is a novel approach that effectively combines convolutional neural network (CNN) and transformer for 3D medical image segmentation. Although the model is based on the 3D U-Net architecture and utilizes the same pre-processing and post-processing methods as nnU-Net, it introduces an additional deformable transformer (DeTrans) encoder. This encoder is designed to capture deeper features from the feature maps produced by the CNN encoder, allowing for better representation of complex image patterns. By bridging the gap between CNNs and transformers, CoTr offers improved performance and greater flexibility in 3D medical image segmentation tasks. The results of the study demonstrate that CoTr outperforms several state-of-the-art models in terms of segmentation accuracy, making it a promising direction for future research in this field.

The nnFormer, proposed by Zhou et al. \cite{23}, is a recent addition to the class of CNN-Transformer hybrid models that has been designed as an extension of the nnUNet architecture. The model consists of several components, including a convolutional embedding layer, a Multi-head Self-attention encoder, and a symmetrical transformer block with a deconvolution-based upsampling layer as the decoder. The convolutional embedding layer maps the input image to a feature space, which is then processed by the Multi-head Self-attention encoder to capture global contextual information. The symmetrical transformer block is responsible for capturing long-range dependencies and generating highly contextualized features. Finally, the deconvolution-based upsampling layer is used to reconstruct the output mask. This model has been shown to be highly effective in 3D medical image segmentation tasks and outperforms several state-of-the-art methods.

The paper "Convolution-Free Medical Image Segmentation using Transformers" by Karimi et al \cite{30}. presents a convolutional-free architecture for medical image segmentation using transformers. However, this approach cannot been used for thin segmentation using thick annotations, and the lack of published code and dataset raises concerns about the validity of the reported results.

Recently, the Mamba model \cite{zhang2024infinimotion,zhang2024kmm,zhang2024motion} has demonstrated state-of-the-art efficiency in sequence modeling by leveraging selective state-space layers. Its strong long-range dependency modeling makes it well-suited for medical imaging segmentation, enabling frameworks like UMamba and SegMamba to achieve improved feature extraction and boundary delineation.

In general, the related works mentioned above primarily utilize either convolutional architecture or incorporate transformer layers as part of the encoder-decoder architecture. In contrast, our proposed model deviates significantly from this approach by introducing a convolutional-free architecture. This novel approach aims to explore the potential of non-convolutional architectures for medical image segmentation, potentially enabling the development of more efficient and effective models in this domain.

\subsection{Research Significants}

The field of 3D medical imaging segmentation has significant implications for real-world medical applications. The development of automatic segmentation using computer vision technology \cite{zhang2024motionavatar} can aid radiologists in detecting the focus of the disease and labeling annotations, reducing the time and effort required for manual segmentation. Additionally, segment masks can provide more accurate attributes of diseases, such as the estimation of the volume of brain hemorrhages, aiding in diagnosis and treatment planning. Moreover, the use of segmentation in longitudinal analysis can assist radiologists in predicting the prognosis of certain diseases, such as cancer, and prioritize urgent cases accordingly. The ability to automatically segment 3D medical images has the potential to greatly enhance the accuracy and efficiency of medical diagnosis and treatment, and ultimately improve patient outcomes.

Our proposed convolutional-free method has two significant contributions that outperform existing segmentation models. Firstly, the transformer architecture in our encoder-decoder model can capture global context information better than CNN-based models. The self-attention mechanism in the transformer architecture allows for the model to attend to important regions in the input image, which results in more accurate semantic labeling and generates high-quality masks. Additionally, transformers can capture long-range dependencies in the image, allowing for better modeling of complex relationships between pixels.

Secondly, our method includes domain adaptation between thin slices and thick annotations. This allows us to segment thin slices with high z-resolution and generate an isomorphic mask that accurately represents the volume of focus. This mask can be used in surgical planning and navigation, providing a more precise pre-surgical simulation and direct guidance during surgery. By accurately representing the volume of focus, our method has the potential to improve patient outcomes and reduce surgical complications. Overall, our proposed method represents a significant improvement over existing segmentation models, with potential for real-world impact in medical applications.

\section{Methodology}

\begin{figure}
    \centering
    \includegraphics[width=1\linewidth]{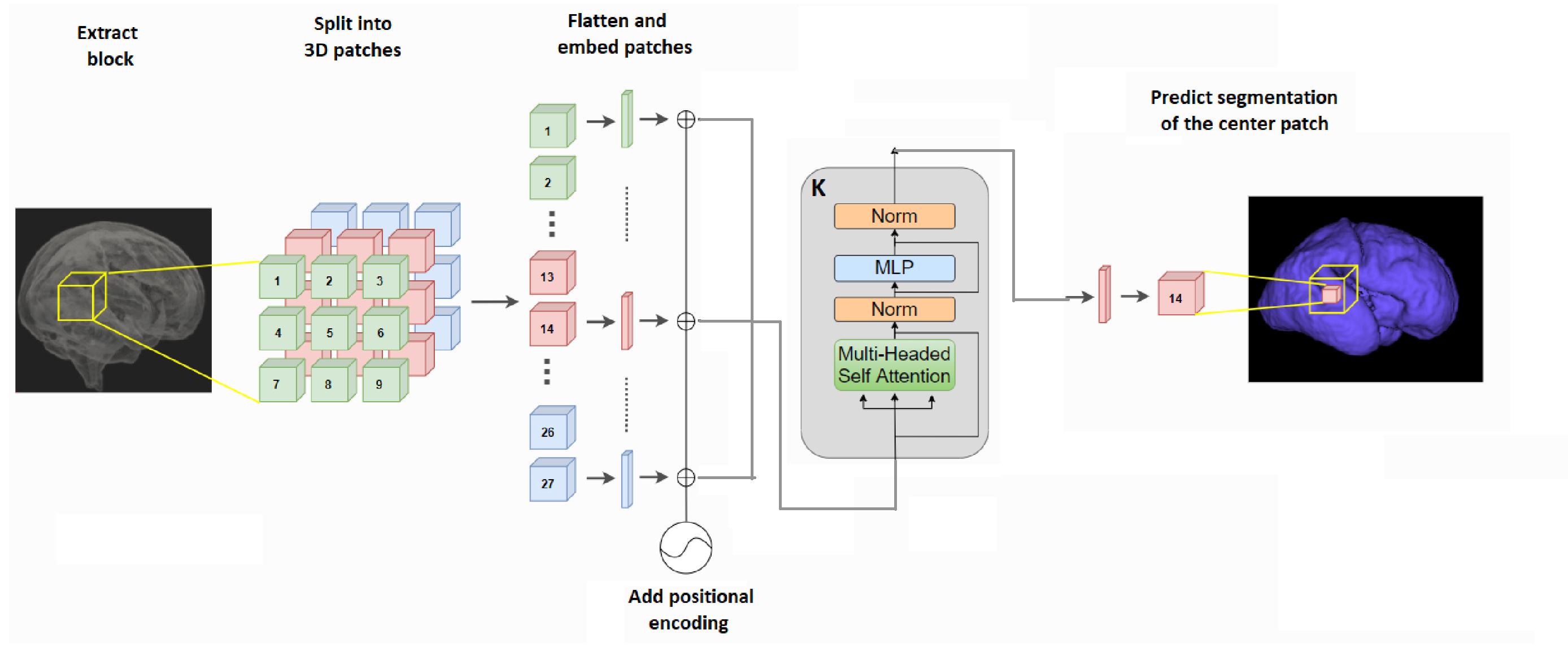} 
    \caption{Proposed pipeline}
    \label{fig:2}
\end{figure}

Our proposed methodology comprises of four key components: an extract block, 3D patch embeddings, 3D patch encodings, and a Transformer encoder coupled with an MLP decoder, which shows in figure \ref{fig:2}

\subsection{Extract Block}

Suppose the shape of the input 3D images is $(channel,height,width,depth)$, we first extract the image into block,

\begin{equation}
  B \in \mathbb{R}^{W \times W \times W \times c}
  \label{eq:1}
\end{equation}

where  $W$ denotes the resolution of the block $B$, and $c$ represents the channel of images.

\subsection{Patch Embedding}

The block $B$ is subdivided into a finite number $ N = n^3$ of contiguous non-overlapping 3D patches,

\begin{equation}
  \{p_i \in \mathbb{R}^{w \times w \times w \times c}\}_{i=1}^N
  \label{eq:2}
\end{equation}

where $n$ represents the number of patches along each dimension of the block, and $ w = W/n $ stands for the resolution of each 3D patch.

The resulting patches are arranged in a specific order to form a sequence. Subsequently, each patch is flattened into a vector of length of $\mathbb{R}^{w^3c}$ and passed through a trainable embedding layer, $E \in \mathbb{R}^{D \times w^3c}$, to yield an embedded representation $E_P \in \mathbb{R}^{D}$.

Note that patches are formed into a sequence, so we can obtained an embedded representation matrix $[E_{P_1};...;E_{P_N}]$ based on the sequence. 

\subsection{Patch Encoding}

We can generate a patch encoding matrix $E_{pos} \in \mathbb{R}^{D \times N}$ using the sine and cosine position encoding functions which demonstrated in Transformer \cite{16}.

Later, we obtained the sequence after patch encoding:

\begin{equation}
  X^0 = [E_{P_1};...;E_{P_N}] + E_{pos}
  \label{eq:3}
\end{equation}

\subsection{Transformer Encoder}

The encoder is composed of a stack of $k$ identical layers (building block), and each building block has two sub-layers:

The first sub-layer is a multi-head self-attention layer consists of $h$ self-attention heads, where the query, key, and value matrices are represented as:

\begin{equation}
  Q = E_QX^k, K = E_KX^k, V = E_VX^k
  \label{eq:4}
\end{equation}

where $E_Q, E_K, E_V \in \mathbb{R}^{D_h \times D}$

Then the attention $A$ of self-attention head is calculated as:

\begin{equation}
  A(Q,K,V) = softmax(\frac{QK^T}{\sqrt{D_h}})V
  \label{eq:5}
\end{equation}

Consequently, the attention $MA$ of multi-head attention layer is calculated as:

\begin{equation}
  MA(Q,K,V) = Concat(Head_1,...,Head_h)W^O
  \label{eq:6}
\end{equation}

where

\begin{equation}
  Head_i = A(QW^Q_i,KW^K_i,VW^V_i)
  \label{eq:7}
\end{equation}

where projections are parameter matrices $W^Q_i, W^K_i, W^V_i$, and $W^O$.

The second sub-layer in the building block is fully connected feed-forward neural network (FFNN), which consists of two linear transformations with a ReLU activation in between.

\begin{equation}
  FFNN(x) = ReLU(xW_1 + b_1)W_2 + b_2
  \label{eq:8}
\end{equation}

The two sub-layer is connected via residual connection, followed by layer normalization. 

\begin{equation}
  Output_{Sublayer} = LayerNorm(x + Sublayer(x))
  \label{eq:9}
\end{equation}

\subsection{MLP Decoder}

The output of the Transformer encoder, $X^k$, will pass through a fully connected linear transformation layer to unify the feature maps into correct channel dimension, and predict mask through a Softmax function for the center patch of the block $B$.

\begin{equation}
  Mask = Softmax(X^kW + b)
  \label{eq:10}
\end{equation}

\subsection{Thin-thick adaptation}

We design a joint weighted loss for the domain adaptation of thin-thick scans, i.e. doing thin-slice segmentation using thick-annotations.

We first generate corresponding thick slices and their ground-truth annotations using given thin slices. The thick slices are obtained from average intensity projection (AIP) \cite{31} and thick anotations are labeled by radiologist. The joint weighted loss for back propagation will be:

\begin{itemize}
    \item Using the ground truth of thick slices for direct supervision, i.e. the prediction of the mask on the thick slices should be the exact ground truth.
    \item The average of the mask intensity of corresponding thin slices should be the ground truth of thick slices.
    \item The average of the feature maps of corresponding thin slices should be the feature map of thick slices.
\end{itemize}

\section{Evaluation Criteria}

\subsection{Evaluation Datasets}

The dataset used for evaluation will be categorized into two categories: (1) public datasets with thick slices and annotations, and (2) a private dataset comprising thin slices and corresponding thick slices and annotations. Notably, there is a lack of publicly available datasets with thin slices and ground truth annotations at present, necessitating the use of a private dataset.

In regards to the thick slice dataset, several publicly available datasets with thick annotations are available. One prominent example is the Medical Segmentation Decathlon \cite{32}, which offers a generalizable 3D semantic segmentation dataset consisting of various medical conditions such as liver tumors, brain tumors, lung tumors, colon cancer, pancreas tumors, and other similar conditions, totaling up to 10 different datasets.

Regarding the private dataset, we specifically aim to collect brain hemorrhage data as this condition is commonly scanned using thin slices. The dataset is also multi-semantic with five distinct labels, namely epidural hemorrhage (EDH), intracerebral hemorrhage (ICH), intraventricular hemorrhage (IVH), subarachnoid hemorrhage (SAH), and subdural hemorrhage (SDH) \cite{33}, making it a challenging segmentation task in the field of medical imaging analysis. The absence of publicly available datasets containing thin slices and corresponding ground truth annotations makes it imperative for us to gather our own data. Additionally, this dataset will be publicly released to address the current lack of public thin-slice data for medical imaging segmentation, and will serve as a new benchmark for multi-semantic segmentation in medical imaging.

\subsection{Experiment Design}

The experimental evaluation of the proposed model will be conducted in two phases: a comparative experiment and an ablation experiment. 

The comparative experiment will compare the performance of different 3D medical imaging segmentation models, including our proposed method, from two perspectives: thick segmentation based on thick annotations using public datasets, and thin segmentation based on thin annotations using private datasets. 

In the subsequent ablation experiment, a systematic analysis of the proposed method's individual components will be conducted to assess its contributions to the field. This will allow us to isolate each variation of the proposed method and provide evidence of their respective impacts, thus contributing to the broader medical imaging segmentation research community.

\subsection{Evaluation Matrices}

In the context of 3D semantic segmentation in medical imaging, two prominent evaluation metrics are used, namely mean Intersection over Union (mIoU) \cite{34} and Dice Similarity Coefficient (DSC) \cite{35}. The former is a measure of the intersection over union between the predicted and ground-truth segmentation masks, while the latter is a measure of the similarity between the predicted and ground-truth masks. These metrics are commonly employed to assess the accuracy and quality of segmentation results in medical imaging, and will be used in our proposed model evaluation as well.

\subsubsection{mIoU}

The mean Intersection over Union (mIoU) is a commonly used evaluation metric in general computer vision segmentation tasks. It measures the similarity between the predicted segmentation mask and the ground truth mask by calculating the intersection over union (IoU) for each class and then taking the average across all classes. The IoU is calculated as the ratio of the intersection between the predicted and ground truth mask to the union of both masks. The mIoU ranges from 0 to 1, where a value of 1 indicates a perfect segmentation and a value of 0 indicates no overlap between the predicted and ground truth masks.

\begin{equation}
  Intersection = GroundTruth \cap Prediction
  \label{eq:11}
\end{equation}

\begin{equation}
  Union = GroundTruth \cup Prediction
  \label{eq:12}
\end{equation}

\begin{equation}
  IoU = Intersection / Union
  \label{eq:13}
\end{equation}

\begin{equation}
  mIoU = mean(IoU_1, ...IoU_n)
  \label{eq:14}
\end{equation}

where $n$ is the number of semantic labels.

\subsubsection{DSC}

Dice Similarity Coefficient (DSC) is a widely used evaluation metric in medical imaging segmentation tasks. It measures the overlap between the predicted segmentation and the ground truth segmentation. The DSC ranges from 0 to 1, where 0 indicates no overlap and 1 indicates a perfect match between the predicted and ground truth segmentations. 

\begin{equation}
  DSC = (2 \times Intersection)/Union
  \label{eq:15}
\end{equation}

Where $Intersection$ represents the intersection of the predicted and ground truth segmentation, and $Union$ represents the Area of the predicted segmentation + Area of the ground truth segmentation.

It is a useful metric for evaluating the accuracy of segmentation models, especially in scenarios where the class imbalance is significant.

\section{Contributions}

\textbf{1: Data acquisition}

\begin{itemize}
    \item Obtain thin-slice data and generate thick slice data based on thin slices
    \item Annotate the thin slice data for multi-semantic segmentation on brain hemorrhage
\end{itemize}

\textbf{2: Model development}

\begin{itemize}
    \item Implement the multi-semantic segmentation model using transformers
    \item Implement the backpropagation of loss for thin-thick adaptation
\end{itemize}

\textbf{3: Experimentation and analysis}

\begin{itemize}
    \item Run comparative experiments on both public thick slice datasets and private thin slice datasets using existing models for comparison
    \item Analyze and evaluate the performance of the model in comparison to existing models
\end{itemize}

\textbf{4: Ablation studies}

\begin{itemize}
    \item Perform ablation studies to demonstrate the contribution of the proposed novel model
\end{itemize}

\textbf{5: Paper writing and review}

\begin{itemize}
    \item Write the paper detailing the work, methodology, results, and findings
    \item Submit the paper to the supervisor for review and feedback
    \item Revise the paper based on supervisor feedback and finalize the paper for submission
\end{itemize}

\section{Limitations and Potential Problems}

One potential limitation in 3D multi-semantic medical imaging segmentation is the long-tail problem \cite{36,37}, which refers to an extreme scenario of unbalanced data in machine learning classification tasks \cite{zhang2024jointvit,qi2025medconv}, especially for multi-label and multi-class classification \cite{wu2024xlip}. An ideal training dataset for a deep learning model should follow a universal distribution, rather than a normal distribution. In the case of mildly unbalanced data, random downsampling \cite{38} of the majority class can be used to balance the data. However, for cases where the gap between the minority and majority class is too large for downsampling, a weighted loss \cite{39} can be designed for backpropagation, where more weight is assigned to the minority class. Despite these approaches, long-tail problems remain challenging. However, the recent development of DiffuMask \cite{40}, a diffusion model for generating mask-image pairs, offers a potential solution for generating long-tail classes and can be employed in medical imaging segmentation for pre-training purposes.

The second potential problem in 3D multi-semantic medical imaging segmentation pertains to the inadequacy of pre-trained models in this domain as compared to general computer vision. Despite a vast collection of unlabeled images in public datasets, the unavailability of suitable pre-trained models poses a challenge. To overcome this, there is a huge potential for developing large-scale self-supervised or contrastive pre-training models based on existing techniques like MoCo \cite{42} or MAE \cite{43}. The pre-trained encoder obtained through this approach can be frozen and leveraged as a feature extractor to achieve effective medical image segmentation.

The third potential limitation and problem in 3D multi-semantic medical imaging segmentation is the ethical and security concerns surrounding data acquisition. Medical imaging data contains sensitive personal information of patients, and therefore, it is crucial to ensure the privacy and confidentiality of the patients' data. The medical community must adhere to strict ethical guidelines and regulations when collecting, storing, and sharing medical data.

One challenge in this area is determining what kind of data can be shared without compromising patients' privacy. There is a need for clear guidelines and policies on the types of medical imaging data that can be shared publicly or among researchers, and under what conditions. Moreover, since medical imaging datasets can be massive, there is a risk of data breaches and hacking, which can lead to the disclosure of patients' sensitive information.

To address these concerns, researchers should take extra precautions to ensure the security and privacy of medical imaging data, such as using secure storage systems, implementing data access controls, and anonymizing patient data. Additionally, obtaining informed consent from patients is essential before collecting and using their medical data for research purposes. By prioritizing the ethical and security concerns surrounding medical imaging data acquisition, researchers can ensure that they are using patient data in a responsible and transparent manner.

\section{Conclusions}

The aim of our research was to improve the accuracy of semantic segmentation on 3D medical images by developing a novel and efficient model that addresses the domain adaptation problem between thick and thin slices. To this end, we proposed a fully convolutional-free architecture based on transformers that can effectively capture global content of images and generate accurate semantic labels and segmentation masks. Our model employs self-attention mechanisms and other techniques to enhance feature extraction and decoding of masks, which makes it more efficient than existing convolutional-based models. Additionally, we designed a joint loss function that supervises the model to segment thin slices based on thick annotations, which improves segmentation accuracy. We also contributed a new benchmark dataset of thin slice multi-semantic segmentation, which is a valuable resource for researchers to evaluate and compare their models. These contributions are significant in advancing the state-of-the-art in medical image analysis and improving diagnostic medicine.

\section{Acknowledgements}
I would like to acknowledge the support of Zeyu Zhang and AI Geeks in contributing to this research project.

% \bibliographystyle{plain}
% \bibliography{main}

\begin{thebibliography}{10}

\bibitem{26}
Boxwala AA and Rosenman JG.
\newblock Retrospective reconstruction of three-dimensional radiotherapy treatment plans of the thorax from two dimensional planning data.
\newblock {\em Int J Radiat Oncol Biol Phys.}, 28(4):1009--15, 1994.

\bibitem{32}
M.~Antonelli, A.~Reinke, S.~Bakas, and et~al.
\newblock The medical segmentation decathlon.
\newblock {\em Nat Commun}, 13(4128), 2022.

\bibitem{8}
V.~Badrinarayanan, A.~Kendall, and R.~Cipolla.
\newblock Segnet: A deep convolutional encoder-decoder architecture for image segmentation.
\newblock {\em IEEE Transactions on Pattern Analysis and Machine Intelligence}, 39(12):2481--2495, 2017.

\bibitem{cai2024medical}
Guohui Cai, Ying Cai, Zeyu Zhang, Yuanzhouhan Cao, Lin Wu, Daji Ergu, Zhinbin Liao, and Yang Zhao.
\newblock Medical ai for early detection of lung cancer: A survey.
\newblock {\em arXiv preprint arXiv:2410.14769}, 2024.

\bibitem{cai2024msdet}
Guohui Cai, Ying Cai, Zeyu Zhang, Daji Ergu, Yuanzhouhan Cao, Binbin Hu, Zhibin Liao, and Yang Zhao.
\newblock Msdet: Receptive field enhanced multiscale detection for tiny pulmonary nodule.
\newblock {\em arXiv preprint arXiv:2409.14028}, 2024.

\bibitem{19}
Nicolas Carion, Francisco Massa, Gabriel Synnaeve, Nicolas Usunier, Alexander Kirillov, and Sergey Zagoruyko.
\newblock End-to-end object detection with transformers.
\newblock In {\em ECCV}, 2020.

\bibitem{21}
Jieneng Chen, Yongyi Lu, Qihang Yu, Xiangde Luo, Ehsan Adeli, Yan Wang, Le~Lu, Alan~L. Yuille, and Yuyin Zhou.
\newblock Transunet: Transformers make strong encoders for medical image segmentation.
\newblock {\em CoRR}, abs/2102.04306, 2021.

\bibitem{20}
Bowen Cheng, Ishan Misra, Alexander~G. Schwing, Alexander Kirillov, and Rohit Girdhar.
\newblock Masked-attention mask transformer for universal image segmentation.
\newblock In {\em CVPR}, 2022.

\bibitem{9}
J.~Deng, W.~Dong, R.~Socher, L.~J. Li, Kai Li, and Li~Fei-Fei.
\newblock Imagenet: A large-scale hierarchical image database.
\newblock In {\em CVPR 2009}, 2009.

\bibitem{15}
Alexey Dosovitskiy, Lucas Beyer, Alexander Kolesnikov, et~al.
\newblock An image is worth 16x16 words: Transformers for image recognition at scale.
\newblock In {\em ICLR}, 2021.

\bibitem{39}
K.~R.~M. Fernando and C.~P. Tsokos.
\newblock Dynamically weighted balanced loss: Class imbalanced learning and confidence calibration of deep neural networks.
\newblock {\em IEEE Transactions on Neural Networks and Learning}, 33(7):2940--2951, 2022.

\bibitem{40}
K.~R.~M. Fernando and C.~P. Tsokos.
\newblock Dynamically weighted balanced loss: Class imbalanced learning and confidence calibration of deep neural networks.
\newblock {\em IEEE Transactions on Neural Networks and Learning}, 33(7):2940--2951, 2022.

\bibitem{ge2024esa}
Jinchao Ge, Zeyu Zhang, Minh~Hieu Phan, Bowen Zhang, Akide Liu, and Yang Zhao.
\newblock Esa: Annotation-efficient active learning for semantic segmentation.
\newblock {\em arXiv preprint arXiv:2408.13491}, 2024.

\bibitem{24}
Radhiana H, Syazarina SO, Shahizon~Azura MM, Hilwati H, and Sobri MA.
\newblock Non-contrast computed tomography in acute ischaemic stroke: A pictorial review.
\newblock {\em Med J Malaysia}, 68(1):93--100, 2013.

\bibitem{43}
Kaiming He, Xinlei Chen, Saining Xie, Yanghao Li, Piotr Dollár, and Ross Girshick.
\newblock Masked autoencoders are scalable vision learners.
\newblock In {\em CVPR}, 2022.

\bibitem{42}
Kaiming He, Haoqi Fan, Yuxin Wu, Saining Xie, and Ross Girshick.
\newblock Momentum contrast for unsupervised visual representation learning.
\newblock In {\em CVPR}, 2020.

\bibitem{6}
Kaiming He, Xiangyu Zhang, Shaoqing Ren, and Jian Sun.
\newblock Deep residual learning for image recognition.
\newblock In {\em CVPR 2016}, 2016.

\bibitem{hiwase2025can}
Abhiram~D Hiwase, Christopher~D Ovenden, Lola~M Kaukas, Mark Finnis, Zeyu Zhang, Stephanie O'Connor, Ngee Foo, Benjamin Reddi, Adam~J Wells, and Daniel~Y Ellis.
\newblock Can rotational thromboelastometry rapidly identify theragnostic targets in isolated traumatic brain injury?
\newblock {\em Emergency Medicine Australasia}, 37(1):e14480, 2025.

\bibitem{14}
Fabian Isensee, Paul~F. Jaeger, Simon A.~A. Kohl, and Klaus~H. Maier-Hein.
\newblock nnu-net: a self-configuring method for deep learning-based biomedical image segmentation.
\newblock {\em Nature Methods}, 18:203--211, 2021.

\bibitem{ji2024sine}
Yiping Ji, Hemanth Saratchandran, Cameron Gordon, Zeyu Zhang, and Simon Lucey.
\newblock Sine activated low-rank matrices for parameter efficient learning.
\newblock {\em arXiv preprint arXiv:2403.19243}, 2024.

\bibitem{25}
Huang K, Rhee DJ, Ger R, Layman R, Yang J, Cardenas CE, and Court LE.
\newblock Impact of slice thickness, pixel size, and ct dose on the performance of automatic contouring algorithms.
\newblock {\em Appl Clin Med Phys.}, 22(5):168--174, 2021.

\bibitem{30}
Davood Karimi, Serge~Didenko Vasylechko, and Ali Gholipour.
\newblock Convolution-free medical image segmentation using transformer networks.
\newblock In {\em MICCAI 2021}, 2021.

\bibitem{10}
Eric Kauderer-Abrams.
\newblock Quantifying translation-invariance in convolutional neural networks.
\newblock {\em CoRR}, abs/1801.01450, 2018.

\bibitem{4}
Alex Krizhevsky, Ilya Sutskever, and Geoffrey~E. Hinton.
\newblock Imagenet classification with deep convolutional neural networks.
\newblock In {\em NIPS 2012}, 2012.

\bibitem{2}
Matthew Lai.
\newblock Deep learning for medical image segmentation.
\newblock {\em CoRR}, abs/1505.02000, 2015.

\bibitem{3}
Y.~Lecun, L.~Bottou, Y.~Bengio, and P.~Haffner.
\newblock Gradient-based learning applied to document recognition.
\newblock {\em Proceedings of the IEEE}, 86(11):2278--2324, 1998.

\bibitem{29}
Kisuk Lee, Jonathan Zung, Peter Li, Viren Jain, and H.~Sebastian Seung.
\newblock Superhuman accuracy on the snemi3d connectomics challenge.
\newblock {\em CoRR}, abs/1706.00120, 2017.

\bibitem{38}
Wonjae Lee and Kangwon Seo.
\newblock Downsampling for binary classification with a highly imbalanced dataset using active learning.
\newblock {\em Big Data Research}, 28:100314, 2022.

\bibitem{18}
Ze~Liu, Yutong Lin, Yue Cao, Han Hu, Yixuan Wei, Zheng Zhang, Stephen Lin, and Baining Guo.
\newblock Swin transformer: Hierarchical vision transformer using shifted windows.
\newblock In {\em ICCV}, 2021.

\bibitem{28}
Fausto Milletari, Seyed-Ahmad Ahmadi, Christine Kroll, and Annika~Plate et~al.
\newblock Hough-cnn: Deep learning for segmentation of deep brain regions in mri and ultrasound.
\newblock {\em Computer Vision and Image Understanding}, 164:92--102, 2017.

\bibitem{qi2025projectedex}
Xuyin Qi, Zeyu Zhang, Aaron~Berliano Handoko, Huazhan Zheng, Mingxi Chen, Ta~Duc Huy, Vu~Minh~Hieu Phan, Lei Zhang, Linqi Cheng, Shiyu Jiang, et~al.
\newblock Projectedex: Enhancing generation in explainable ai for prostate cancer.
\newblock {\em arXiv preprint arXiv:2501.01392}, 2025.

\bibitem{qi2025medconv}
Xuyin Qi, Zeyu Zhang, Huazhan Zheng, Mingxi Chen, Numan Kutaiba, Ruth Lim, Cherie Chiang, Zi~En Tham, Xuan Ren, Wenxin Zhang, et~al.
\newblock Medconv: Convolutions beat transformers on long-tailed bone density prediction.
\newblock {\em arXiv preprint arXiv:2502.00631}, 2025.

\bibitem{17}
Prajit Ramachandran, Niki Parmar, Ashish Vaswani, Irwan Bello, Anselm Levskaya, and Jon Shlens.
\newblock Stand-alone self-attention in vision models.
\newblock In {\em NIPS}, 2019.

\bibitem{33}
C.S. Rau, S.C. Wu, S.Y. Hsu, H.T. Liu, C.Y. Huang, T.M. Hsieh, S.E. Chou, W.T. Su, Y.W. Liu, and C.H. Hsieh.
\newblock Concurrent types of intracranial hemorrhage are associated with a higher mortality rate in adult patients with traumatic subarachnoid hemorrhage: A cross-sectional retrospective study.
\newblock {\em Int J Environ Res Public Health}, 16(23):4787, 2019.

\bibitem{7}
J.~Redmon, S.~Divvala, R.~Girshick, and A.~Farhadi.
\newblock You only look once: Unified, real-time object detection.
\newblock In {\em CVPR 2016}, 2016.

\bibitem{11}
Olaf Ronneberger, Philipp Fischer, and Thomas Brox.
\newblock U-net: Convolutional networks for biomedical image segmentation.
\newblock In {\em MICCAI}, 2015.

\bibitem{27}
L.~Shen, W.~Zhao, and L.~Xing.
\newblock Patient-specific reconstruction of volumetric computed tomography images from a single projection view via deep learning.
\newblock {\em Nat Biomed Eng}, 3:880--888, 2019.

\bibitem{31}
K.~Shirai, K.~Nishiyama, T.~Katsuda, Y.~Ueda, M.~Miyazaki, K.~Tsujii, and S.~Ueyama.
\newblock Maximum intensity projection (mip) and average intensity projection (aip) in image guided stereotactic body radiation therapy (sbrt) for lung cancer.
\newblock {\em International Journal of Radiation Oncology, Biology, Physics}, 84(3), 2012.

\bibitem{5}
Karen Simonyan and Andrew Zisserman.
\newblock Very deep convolutional networks for large-scale image recognition.
\newblock In {\em ICLR 2015}, 2015.

\bibitem{tan2024segkan}
Shengbo Tan, Rundong Xue, Shipeng Luo, Zeyu Zhang, Xinran Wang, Lei Zhang, Daji Ergu, Zhang Yi, Yang Zhao, and Ying Cai.
\newblock Segkan: High-resolution medical image segmentation with long-distance dependencies.
\newblock {\em arXiv preprint arXiv:2412.19990}, 2024.

\bibitem{tan2024segstitch}
Shengbo Tan, Zeyu Zhang, Ying Cai, Daji Ergu, Lin Wu, Binbin Hu, Pengzhang Yu, and Yang Zhao.
\newblock Segstitch: Multidimensional transformer for robust and efficient medical imaging segmentation.
\newblock {\em arXiv preprint arXiv:2408.00496}, 2024.

\bibitem{34}
Vlad Taran, Nikita Gordienko, Yuriy Kochura, Yuri Gordienko, Alexandr Rokovyi, Oleg Alienin, and Sergii Stirenko.
\newblock Performance evaluation of deep learning networks for semantic segmentation of traffic stereo-pair images.
\newblock In {\em International Conference on Computer Systems and Technologies 2018}, 2018.

\bibitem{1}
Martin Thoma.
\newblock A survey of semantic segmentation.
\newblock {\em CoRR}, abs/1602.06541, 2016.

\bibitem{16}
Ashish Vaswani, Noam Shazeer, Niki Parmar, Jakob Uszkoreit, et~al.
\newblock Attention is all you need.
\newblock In {\em NIPS}, 2017.

\bibitem{36}
Yu-Xiong Wang, Deva Ramanan, and Martial Hebert.
\newblock Learning to model the tail.
\newblock In {\em NIPS}, 2017.

\bibitem{wu2023bhsd}
Biao Wu, Yutong Xie, Zeyu Zhang, Jinchao Ge, Kaspar Yaxley, Suzan Bahadir, Qi~Wu, Yifan Liu, and Minh-Son To.
\newblock Bhsd: A 3d multi-class brain hemorrhage segmentation dataset.
\newblock In {\em International Workshop on Machine Learning in Medical Imaging}, pages 147--156. Springer, 2023.

\bibitem{wu2024xlip}
Biao Wu, Yutong Xie, Zeyu Zhang, Minh~Hieu Phan, Qi~Chen, Ling Chen, and Qi~Wu.
\newblock Xlip: Cross-modal attention masked modelling for medical language-image pre-training.
\newblock {\em arXiv preprint arXiv:2407.19546}, 2024.

\bibitem{zhang2025gamed}
Ruicheng Zhang, Haowei Guo, Zeyu Zhang, Puxin Yan, and Shen Zhao.
\newblock Gamed-snake: Gradient-aware adaptive momentum evolution deep snake model for multi-organ segmentation.
\newblock {\em arXiv preprint arXiv:2501.12844}, 2025.

\bibitem{37}
Yifan Zhang, Bingyi Kang, Bryan Hooi, Shuicheng Yan, and Jiashi Feng.
\newblock Deep long-tailed learning: A survey.
\newblock {\em CoRR}, abs/2110.04596, 2021.

\bibitem{zhang2024deep}
Zeyu Zhang, Khandaker~Asif Ahmed, Md~Rakibul Hasan, Tom Gedeon, and Md~Zakir Hossain.
\newblock A deep learning approach to diabetes diagnosis.
\newblock In {\em Asian Conference on Intelligent Information and Database Systems}, pages 87--99. Springer, 2024.

\bibitem{zhang2024kmm}
Zeyu Zhang, Hang Gao, Akide Liu, Qi~Chen, Feng Chen, Yiran Wang, Danning Li, and Hao Tang.
\newblock Kmm: Key frame mask mamba for extended motion generation.
\newblock {\em arXiv preprint arXiv:2411.06481}, 2024.

\bibitem{zhang2024infinimotion}
Zeyu Zhang, Akide Liu, Qi~Chen, Feng Chen, Ian Reid, Richard Hartley, Bohan Zhuang, and Hao Tang.
\newblock Infinimotion: Mamba boosts memory in transformer for arbitrary long motion generation.
\newblock {\em arXiv preprint arXiv:2407.10061}, 2024.

\bibitem{zhang2024motion}
Zeyu Zhang, Akide Liu, Ian Reid, Richard Hartley, Bohan Zhuang, and Hao Tang.
\newblock Motion mamba: Efficient and long sequence motion generation.
\newblock In {\em European Conference on Computer Vision}, pages 265--282. Springer, 2024.

\bibitem{zhang2024jointvit}
Zeyu Zhang, Xuyin Qi, Mingxi Chen, Guangxi Li, Ryan Pham, Ayub Qassim, Ella Berry, Zhibin Liao, Owen Siggs, Robert Mclaughlin, et~al.
\newblock Jointvit: Modeling oxygen saturation levels with joint supervision on long-tailed octa.
\newblock In {\em Annual Conference on Medical Image Understanding and Analysis}, pages 158--172. Springer, 2024.

\bibitem{zhang2024segreg}
Zeyu Zhang, Xuyin Qi, Bowen Zhang, Biao Wu, Hien Le, Bora Jeong, Zhibin Liao, Yunxiang Liu, Johan Verjans, Minh-Son To, et~al.
\newblock Segreg: Segmenting oars by registering mr images and ct annotations.
\newblock In {\em 2024 IEEE International Symposium on Biomedical Imaging (ISBI)}, pages 1--5. IEEE, 2024.

\bibitem{zhang2024motionavatar}
Zeyu Zhang, Yiran Wang, Biao Wu, Shuo Chen, Zhiyuan Zhang, Shiya Huang, Wenbo Zhang, Meng Fang, Ling Chen, and Yang Zhao.
\newblock Motion avatar: Generate human and animal avatars with arbitrary motion.
\newblock {\em arXiv preprint arXiv:2405.11286}, 2024.

\bibitem{zhang2024meddet}
Zeyu Zhang, Nengmin Yi, Shengbo Tan, Ying Cai, Yi~Yang, Lei Xu, Qingtai Li, Zhang Yi, Daji Ergu, and Yang Zhao.
\newblock Meddet: Generative adversarial distillation for efficient cervical disc herniation detection.
\newblock In {\em 2024 IEEE International Conference on Bioinformatics and Biomedicine (BIBM)}, pages 4024--4027. IEEE, 2024.

\bibitem{zhang2023thinthick}
Zeyu Zhang, Bowen Zhang, Abhiram Hiwase, Christen Barras, Feng Chen, Biao Wu, Adam~James Wells, Daniel~Y Ellis, Benjamin Reddi, Andrew~William Burgan, Minh-Son To, Ian Reid, and Richard Hartley.
\newblock Thin-thick adapter: Segmenting thin scans using thick annotations.
\newblock {\em OpenReview}, 2023.

\bibitem{zhao2024landmark}
Yang Zhao, Zhibin Liao, Yunxiang Liu, Koen~Oude Nijhuis, Britt Barvelink, Jasper Prijs, Joost Colaris, Mathieu Wijffels, Max Reijman, Zeyu Zhang, et~al.
\newblock A landmark-based approach for instability prediction in distal radius fractures.
\newblock In {\em 2024 IEEE International Symposium on Biomedical Imaging (ISBI)}, pages 1--5. IEEE, 2024.

\bibitem{23}
Hong-Yu Zhou, Jiansen Guo, Yinghao Zhang, Lequan Yu, Liansheng Wang, and Yizhou Yu.
\newblock nnformer: Interleaved transformer for volumetric segmentation.
\newblock {\em CoRR}, abs/2109.03201, 2021.

\bibitem{12}
Zongwei Zhou, Md~Mahfuzur Rahman~Siddiquee, Nima Tajbakhsh, and Jianming Liang.
\newblock Unet++: A nested u-net architecture for medical image segmentation.
\newblock In {\em International Workshop on Deep Learning in Medical Image Analysis}, volume 11045, pages 3--11, 2018.

\bibitem{35}
K.H. Zou, S.K. Warfield, A.~Bharatha, C.M. Tempany, M.R. Kaus, S.J. Haker, W.M.~3rd Wells, F.A. Jolesz, and R.~Kikinis.
\newblock Statistical validation of image segmentation quality based on a spatial overlap index.
\newblock {\em Acad Radiol}, 11(2):178--89, 2004.

\bibitem{13}
Özgün Çiçek, Ahmed Abdulkadir, Sonja~S. Lienkamp, Thomas Brox, and Olaf Ronneberger.
\newblock 3d u-net: Learning dense volumetric segmentation from sparse annotation.
\newblock In {\em MICCAI}, 2016.

\end{thebibliography}

\end{document}